# A self-learning magnetic Hopfield neural network with intrinsic gradient descent adaption.

Chang Niu[1,2], Huanyu Zhang[1,2], Chuanlong Xu[1,2], Wenjie Hu[1,2], Yunzhuo Wu[1,2], Yu Wu[1,2], Yadi Wang[1,2], Tong Wu[1,2], Yi Zhu[1,2], Yinyan Zhu[1,4,5], Wenbin Wang[1,3,4,5], Yizheng Wu[1,2], Lifeng Yin[1,2,4,5,6], Jiang Xiao[1,2,3,4,5,6], Weichao Yu[1,5]*, Hangwen Guo[1,3,4,5]† and Jian Shen[1,2,3,4,5,6]‡

1 State Key Laboratory of Surface Physics and Institute for Nanoelectronic Devices and Quantum Computing, Fudan University, Shanghai 200433, China

2 Department of Physics, Fudan University, Shanghai 200433, China

3 Shanghai Branch, Hefei National Laboratory, Shanghai 201315, China

4 Shanghai Research Center for Quantum Sciences, Shanghai 201315, China

5 Zhangjiang Fudan International Innovation Center, Fudan University, Shanghai 201210, China

6 Collaborative Innovation Center of Advanced Microstructures, Nanjing 210093, China

Weichao Yu*, Hangwen Guo† and Jian Shen‡.

**Email:** wcyu@fudan.edu.cn, hangwenguo@fudan.edu.cn, shenj5494@fudan.edu.cn

**This PDF file includes:**

Main Text
Figures 1 to 5
Table 1

## Abstract

Physical neural networks using physical materials and devices to mimic synapses and neurons offer an energy-efficient way to implement artificial neural networks. Yet, training physical neural networks are difficult and heavily relies on external computing resources. An emerging concept to solve this issue is called physical self-learning that uses intrinsic physical parameters as trainable weights. Under external inputs (i.e. training data), training is achieved by the natural evolution of physical parameters that intrinsically adapt modern learning rules via autonomous physical process, eliminating the requirements on external computation resources. Here, we demonstrate a real spintronic system that mimics Hopfield neural networks (HNN) and unsupervised learning is intrinsically performed via the evolution of physical process. Using magnetic texture-defined conductance matrix as trainable weights, we illustrate that under external voltage inputs, the conductance matrix naturally evolves and adapts Oja's learning algorithm in a gradient descent manner. The self-learning HNN is scalable and can achieve associative memories on patterns with high similarities. The fast spin dynamics and reconfigurability of magnetic textures offer an advantageous platform towards efficient autonomous training directly in materials.

**Significance Statement**

Natural physical systems evolve with certain global quantities being minimized or maximized due to physical laws. For example, charges in conductors redistribute to reach electrostatic equilibrium, minimizing electrostatic energy, and gases spread to maximize entropy. This research leverages these natural efficiencies by encoding optimization problems, like training artificial neural networks, into the evolution of physical systems. The concept is called "physical self-learning" where systems' intrinsic parameters autonomously evolve guided by natural laws. Specifically, a physical Hopfield neural network using a magnetic thin film is developed. Inputs are encoded as electric signals that manipulate magnetic textures within the film through Oersted fields, enabling the film to learn from external inputs and perform tasks like associative memory of similar patterns.

**Main Text**

**Introduction**

In the past decade, the field of artificial neural networks has grown exponentially and impacts every perspective of our daily life, yet facing the challenges of huge energy consumption in both training and inference stages. To emulate energy-efficient computational capabilities of biological brains, physical neural networks (PNN) that uses physical devices to mimic synaptic and neuron functionalities has become one of the promising paths. While numerous physical systems (i.e. electronics, photonic, spintronic, ionic, superconducting etc.) have been implemented as PNNs (1–8), so far, most experimental systems have been focusing on the inference phase. In contrast, as illustrated in Figure 1b, training such physical system, a process to fine-tune and optimize synaptic weights, heavily relies on the iterative feedback from external computers or circuits, posing a bottleneck on its training efficiency. Recently, efforts have been focused on the learning/training phase with various type of experimental systems, e.g., training deep PNN with (9) or without (10) backpropagation and fully forward training in optical system (11). Although external computation is still required, these strategies may facilitate learning faster and more energy-efficiently than conventional electronic processors. In this work, we focus on an emerging solution called physical self-learning where a PNN can be trained by autonomous and internal physical processes, as depicted in Fig.1c (12–15). During physical self-learning, the training data are encoded as external inputs and the trainable weights are mapped as tunable physical parameters (12, 14, 16–20). Under external inputs, the physical parameters autonomously evolve towards the direction that the total energy of the physical system decreases through a dissipative process. For instance, optimization can be achieved using memristive circuits exhibiting asymptotic behavior (21–23). Similarly, the concept of equilibrium propagation, a two-phase learning scheme that relies on local information, demonstrates how supervised learning can be implemented within physical systems (24). This study primarily explores the application of physical self-learning in the context of an unsupervised learning task. When designed appropriately, such intrinsic physical process can follow certain learning rules which is equivalent to the training algorithms applied in digital computers that minimize a cost function, so that training is accomplished in the physical system without external computations just like biological neurons (Fig.1a). Therefore, finding physical systems whose physical process naturally adapts modern training algorithms can improve training efficiency significantly and offer a vehicle towards the ultimate purpose of "natural intelligence" as in biological systems (25, 26).

A milestone model in the field of machine learning is Hopfield neural network (HNN) (27). When originally proposed, the HNN features fully-connected neurons which is the foundation of modern models such as Boltzmann machines and Ising machines (28, 29). Recently, the importance of HNN has re-emerged to be closely related to transformer networks for advanced natural language processing (30, 31). Several physical systems such as spintronics, memristors and optics have been implemented as HNNs (31–38). A typical functionality of HNN is associative memory where the network is trained to memorize input patterns. The performance of memorization such as recalling accuracy strongly depends on the underlying training algorithms. For most of the physical implementations of HNNs, simple outer-product rule is adopted to train the network in an *ex situ* manner so that only high-orthogonality (that is, low similarity) patterns can be memorized and later recalled in the inference stage.

Here, we demonstrate a real spintronic platform that forms a self-learning HNN which autonomously adapts training algorithm beyond outer-product rule. The HNN is physically constructed by a Py ($Fe_{80}Ni_{20}$) magnetic disk where the synaptic weights are represented by the spin-dependent conductance matrix (Figure 2b). Using such conductance matrix as physical parameters, we experimentally show that magnetic textures can evolve under external inputs and naturally adapt gradient descent learning of Oja's rule – an advanced training algorithm for unsupervised learning. We further demonstrate by simulation that our self-learning magnetic HNN are scalable and shows unique capability to memorize and recall low-orthogonality patterns with high accuracy, which outperforms usual physical HNNs. Our results suggest that the reconfigurability and nanosecond spin dynamics of magnetic textures in spintronic system offer an advantageous platform towards parallel and efficient self-learning directly in materials for future computing.

**Results**

**Device and working principle**

Spintronic devices have emerged as important building blocks to emulate artificial neural networks (4, 39–49). We choose magnetic permalloy (Py) that contains rich tunable magnetic texture configurations as prototype material for this study (50). Figure 2a shows the schematic of a device consisting three material layers grown on Si wafer substrate: bottom Py, middle insulating $Al_2O_3$ and top Au layer (see Materials and Methods of "**sample fabrication**"). As a proof-of-principle demonstration, we apply the 4-node learning scheme typically considered in memristor networks (23). The four nodes on the top Au layer serve as binary neurons where the input training data are represented by the positive/negative pulse-voltage patterns. The conductance matrix elements $G_{ij}$ in the bottom Py layer are analogous to the synaptic weights where its evolution corresponds to the training process. Taking "$+-+-$" input pattern as an example (Figure 2b), positive voltage pulses are applied in node 1 and 3 in the Au layer, forming a current distribution according to Ohm's law. The current in Au layer thus generates an Oersted field distribution and configures the spin orientation, that is, the magnetic texture configuration in the bottom Py layer (Figure 2b middle) (51). The amplitude of magnetic field generated is about 20 Oe which is larger than the coercive field of our Py layer (less than 10 Oe). The dynamical evolution of the working principle is illustrated in Supplementary Video 1. The evolution of conductance matrix $G_{ij}$ between the four electrodes in the magnetic Py layer can be detected due to the anisotropic magnetoresistance (AMR) effect (52, 53), which states that the local resistivity tensor $\hat{\rho}[\mathbf{m}(\mathbf{r}), \mathbf{r}]$ depends on the angle between the detection current flow and the local magnetization $\mathbf{m}(\mathbf{r})$ as illustrated in Figure 2d. This means that the electrical resistance changes depending on how the magnetization is oriented relative to the direction of the current flow. By measuring these resistance changes, we can effectively map the magnetic configurations within the material using electrical signals. This detection current in Py layer is small (500 μA) and will not change the magnetic texture configuration. By measuring the electrical voltages via the combination of electrodes connections in the Py layer, one can directly quantify the conductance matrix $G_{ij}$ experimentally according to Kirchhoff's law $I_i^{Py} = G_{ij}V_j^{Py}$ (see Materials and Methods of "**electrical setup and determining the conductance matrix**"). Therefore, we have mapped the magnetic device with four electrodes into a 4-node HNN and we will focus in the next section on the self-learning process, i.e., the self-adaptive evolution of conductance matrix.

**Self-learning capability with intrinsic gradient descent**

To demonstrate the self-learning capability, we experimentally measure the evolution of $G_{ij}$ under the applied voltage pulses from the top Au layer. Figure 3a illustrates the evolutions of the conductance matrix values when the "$+ + - -$" pulse pattern is applied. Initially, the magnetic texture is configured by applying a 200 µs, 5 V pulse in the Au layer with "+-+-" voltage pattern. According to **"determining the conductance matrix"** in methods, the corresponding conductance matrix values $G_{ij}^{+-+-}$ (in Siemens) are measured and calculated to be

$$\begin{pmatrix} +0.178 & -0.066 & -0.066 & -0.046 \\ -0.066 & +0.171 & -0.041 & -0.065 \\ -0.066 & -0.041 & +0.173 & -0.066 \\ -0.046 & -0.065 & -0.066 & +0.177 \end{pmatrix} \text{ S.} \qquad (1)$$

The conductance matrix is symmetric and $\sum_j G_{ij}=0$, satisfying the Kirchhoff's law. We then apply a series of "$+ + - -$" pulses (3 V, 50 µs) on the top Au layer and the conductance matrix evolution is shown in Figure 3a: Conductance $G_{12}$ and $G_{34}$ increase, $G_{13}$ and $G_{24}$ decrease and finally saturate, while $G_{23}$ and $G_{14}$ remain unchanged. These observations consist well with the physical process of magnetic textures. Taking node 13 and node 24 for example, the increased electrical currents in the Au layer for the two connections should enhance the Oersted magnetic fields to align the spins between 13 and 24 in the magnetic Py layer, resulting the magnitude increase of matrix elements $|G_{13}|$ and $|G_{24}|$ due to the AMR effect. The reversed process occurs for node 12 and node 34 causing $|G_{12}|$ and $|G_{34}|$ to decrease, respectively. For node 14 and node 23, the electric currents are unaltered so that $|G_{14}|$ and $|G_{23}|$ remain unchanged.

We further performed micromagnetic modelling to capture microscopic dynamics of the magnetic textures, e.g., formation and motion of vortex cores as well as other intermediate states of domain walls (see Materials and Methods of "**micromagnetic simulation**"). The numerical results are plotted in Figure 3a along with the experimental data. Though some conductance fluctuations are observed at early stage, possibly due to the competition of spin dynamics, the evolution and convergence of the numerical results follow well with the experimental curves. Snapshots of magnetic configurations during the evolution is shown in Figure 3b. The above observations can be summarized into two key features: (1) the direction of conductance change $\Delta G_{ij}$ is determined by the polarity of $V_i^{Au}V_j^{Au}$ in the Au layer; (2) $\Delta G_{ij}$ will not diverge, but naturally saturate by certain constraint. Therefore, to capture the nature of conductance matrix evolution, we model the evolution of conductance matrix elements by the following equation:

$$G_{ij}(t) = G_{ij}(t-1) + \eta\left[V_i^{Au}V_j^{Au} - 2\alpha_{ij}\big(G_{ij}(t-1) - G_{ij}^{avg}\big)\right], \qquad (2)$$

where $G_{ij}$ is conductance matrix elements; $\eta$ (= $0.0006 S \cdot V^{-2}$) is a parameter governing the speed of evolution; $V_i^{Au}$ is the voltage of the node *i* in the Au layer, with $G_{ij}^{avg} = \frac{G_{ij}^{min}+G_{ij}^{max}}{2}$ defined as the average of maximum and minimum values which can be reached by matrix elements. The coefficient $\alpha_{ij} = \frac{(\Delta V_i^{Au})^2}{4(G_{ij}^{max}-G_{ij}^{min})}$ can be treated as an effective learning rate ($\alpha_{12} = \alpha_{13} = \alpha_{24} = \alpha_{34} = 111 S^{-1} \cdot V^2, \alpha_{14} = \alpha_{23} = 277 S^{-1} \cdot V^2$), which is proportional to the square of applied voltage difference $\Delta V_i^{Au}$. We note that domain wall memory effect is robust in Py nanostructures due to magnetic shape anisotropy (54, 55). Such effect is minimal in our thin film geometry that should not affect its performance.

The calculated results of Eq. (2) are plotted in Figure 3a and follow well with our experimental and simulation data. Intriguingly, the two terms in Eq. (2) not only captures the two key features described above but is equivalent to the Oja's learning rule (56) applied to perform unsupervised learning in software. The essence of Oja's rule is perfectly captured in our magnetic system: The first term represents the increment characterized by usual Hebbian rule, which originates from magnetization evolution driven by Oersted field and AMR effect: the activation of neurons (virtually represented as nodes in Au layer) strengthens the synaptic weights (conductance matrix in Py layer) connecting them and vice versa, in close analogy with Hebbian learning (57). The second term of Eq. (2) plays as a constraint, which applies increment limit to the conductance change. This is natural since conductance change induced by AMR effect should finally reach saturation for all realistic materials. These results suggest that the evolution of the Oersted field driven magnetic texture realizes the Oja's rule through physical self-learning process, and thus eliminating the iterative feedbacks with external computer or circuits.

Interestingly, the evolution of $G_{ij}$ naturally adapt the concept of gradient descent to minimize the energy function of our magnetic system, which is normally known as cost function $C$ in the language of machine learning via $\Delta G_{ij} = -\eta \frac{dC}{dG_{ij}}$. The cost function can be obtained by integrating Eq. (2):

$$C = \sum_{ij, i \neq j} \frac{\left(V_i^{Au} - V_j^{Au}\right)^2 G_{ij}}{2} + \sum_{ij, i \neq j} \alpha_{ij} \left(G_{ij} - G_{ij}^{max}\right)^2 \quad (3)$$

In contrast to conventional cost function that are defined artificially, the two terms in Eq. (3) are both in dimension of power efficiency and have intrinsic physical meanings. The first term in Eq. (3) represents the essential Ohmic power loss[1] due to the conductance changes on the magnetic layer. When the stimulus is present in the Au layer ($V_i^{Au} \neq V_j^{Au}$), the magnitude of conductance $G_{ij}$ increases, which results in higher Ohmic (39) power loss during measurement in the magnetic layer, and at the same time, minimizes the cost function. The second term also has the dimension of power, representing an extra contribution of Ohmic power loss once the synaptic conductance deviates from its maximum value, which applies a global constraint on the cost function. This mechanism of constraint comes naturally from the planar Hall effect induced by AMR, where elements of local conductivity tensor are related as a function of local magnetization as well as current distribution (12, 54). We also note that the cost function in our magnetic system is in high resemblance to the cost function proposed in adaptive flow networks (58) where evolution of conductance minimizes the cost function under constraint (see Materials and Methods of "**self-learning magnetic network versus adaptive flow network**").

The intrinsic gradient descent learning is also illustrated in other case, for example, from "+ + + −" to "+ − + +". During this evolution, both $|G_{14}|$ and $|G_{34}|$ decrease due to enhanced currents in both node 14 and node 34. The reverse process occurs for node 12 and node 23 causing $|G_{12}|$ and $|G_{23}|$ to increase (**SI Appendix, Figure S1**). In short summary, the magnetic textures present a self-learning physical system with intrinsic gradient descent adaption, providing a fast and natural way to train neural network without external assistance.

**Examination of learning performance via inference**

---

[1] The nondiagonal elements of conductance matrix $G_{ij}$ are negative by definition according to Kirchhoff's law. Therefore, the cost function Eq. (3) reaches minimum while the Ohmic power of the realistic physical system reaches maximum during the gradient descent evolution.

A key functionality of HNN is associative memory – where the network is trained to memorize the input data/pattern. During inference stage, such pattern should be correctly recalled when a distorted pattern is given which can be mathematically written by examining the energy criteria:

$$E = - \sum_{i,j} w_{ij} V_i^{\text{Py}} V_j^{\text{Py}}, \quad (4)$$

which implies that for a trained weight matrix $w_{ij}$, the network will adapt its state $V_i^{\text{Py}}$ under iterative updates until its energy Eq. (4) is minimized, corresponding to a stable state (or an attractor) which is exactly the pattern to be recalled. Interestingly, Eq. (4) takes the same form as Eq. (3) after the self-learning process with $G_{ij}$ fixed (see Materials and Methods of "**relation between cost function of magnetic texture and energy function of HNN**"). This indicates that the self-learned magnetic texture should obey the same energy criteria to perform associative memory. Here, $w_{ij}$ and $G_{ij}$ is connected via $w_{ij} = G_{ij} - G_{ij}^{\text{avg}}$ (see Materials and Methods of "**relation between cost function of magnetic texture and energy function of HNN**").

To validate the above energy criteria in our 4-node magnetic film, we examined all 7 independent input patterns that are reconfigurable by applying different voltage patterns on the top Au layer (see Materials and Methods of "**number of independent configurations**"). The results are shown in Figure 4b-g respectively. Take Figure 4a as an example, the binary voltage pattern "$+ + - -$" is applied in the top Au layer and the conductance matrix in the Py layer evolve and saturate. The final conductance values $G_{ij}$ are experimentally measured and the corresponding weight matrix $w_{ij}$ is obtained (**SI Appendix, Figure S2**). Using this weight matrix, all possible combinations of trial states are fed to calculate the energy according to Eq. (4) and plotted in Figure 4a. Clearly, the energy is minimal for the "$+ + - -$" trial state that is identical to the input pattern, obeying the lowest energy criteria of HNN. The same examinations are correctly performed for other six patterns as in Figure 4b-g. In addition, our experimental device shows good stability and repeatability when switching between these configurations (**SI Appendix, Figure S3**). Also, the above energy examination is successfully performed in another sample (**SI Appendix, Figure S4**), suggesting the robust plasticity and generality on the implementation of HNN against device-to-device variations and imperfections. These results fit well with the nature of physical learning where imperfections are included in the response of the real physical systems (17).

Having validated the energy criteria of our 4-node magnetic HNN, the inference of associative memory can be performed following the inference protocol (see Materials and Methods of "**inference process for associative memory and image recognition**") (35). The memorized patterns can be correctly inferred and retrieved using standard asynchronous/synchronous process, typically within several iterations (**SI Appendix, Figure S5**) (12). Interestingly, using multiple magnetic texture configurations, it allows direct inference of larger patterns by dividing the pattern into small 2×2 pieces (see Materials and Methods of "**inference process for associative memory and image recognition**" & **SI Appendix, Figure S6**). As shown in Figure 4h, 3×5 sized image can be correctly inferred and retrieved.

**Scaling up for enhanced performance**

In the above sections, we have shown an experimental proof-of-principle demonstration of a 4-node magnetic HNN that are capable of self-learning with Oja's learning rule. To potentially serve as functional devices, we explored the scalability and the enhanced functionality of our architecture. Below, we perform micromagnetic simulations based on the experimental data and show that our magnetic HNN are scalable and have the unique ability to memorize similar (low-orthogonal) patterns thanks to the adaption of intrinsic Oja's rule as described in Eq. (2). Such ability remains challenging in most physical HNNs due to the use of simple outer-product training algorithm, i.e., $w_{ij} = \frac{1}{N} \sum_{P=1}^{p} X_i^p X_j^p$, which can only memorize patterns with high orthogonality (35, 59–61). Figure 5a shows the schematics of an expanded 35 nodes magnetic HNN. The scaling strategy distributes all the nodes in arrays within a continuous magnetic thin film, so that the magnetic HNN is fully-connected in principle. However, due to the distance-dependence of magnetoresistance, the interaction between nodes decreases with their distance, leading to an effective local connection between nodes. Such locally-connected network may comprise some memory capacity as compared to fully-connected one (60–62), but offers massive parallelism in the training stage where the whole magnetic textures evolves simultaneously when the voltage pulses are applied in the Au nodes. Such parallel self-learning process can significantly save the training time when scaling up. On the other hand, the scaling strategy avoids wiring problems suffered by fully-connected networks and reduces the requirements for precise measurement of AMR.

We first choose three similar letters E, C and P to demonstrate the associative memory abilities. Training is accomplished by encoding the noiseless images into voltages applied on corresponding electrodes in the Au layer. We train the network by inputting the E, C and P in a sequential way with 5 ns duration for each letter, and one epoch refers to a complete cycle of training (15 ns). Then, images from testing set are input to the network, and the network outputs recalled images according to standard inferring process (**SI Appendix, Figure S7**). We calculate the similarity according to the inner product of recalled picture $x'$ (inferred via standard synchronous process) and original pictures $x_i^0$ ($i$=E, C, P) by following $S_i(x', x_i^0) = x' \cdot x_i^0$.

The recalled letter is recognized as the corresponding letter with maximum similarity. Figure 5b and 5c shows the confusion matrix of the obtained recognition results vs. the desired outputs and the evolution of accuracy with increasing training epochs. The inset of Figure 5c represents the real-time attractors preserved by the Py film. The overall accuracy for correctly recalling three similar letters (E, C, P) by using self-learning Oja's rule can reach 97%, outperforming ~33% performed by outer-product rules. We further compare the accuracy between self-learning Oja's rule and outer-product rule when the capacity of memories increases to (EDCP) and (EDCPL). As shown in Figure 5d, while outer-product with low accuracy would only remember one or two attractors for pictures with low orthogonality, our device can keep high accuracy when increasing the number of memory capacity (**SI Appendix, Figure S8**). High recalling accuracy is also achieved in another example of low orthogonality patterns "08965" (**SI Appendix, Figure S9**), suggesting the generality in using our magnetic textures for this task.

**Fully reconfigurable spin logics**

Finally, besides the applications in neuromorphic computing, our self-learned magnetic texture is capable of performing conventional Boolean logics including AND, OR, NAND and NOR gates (see Materials and Methods of "**boolean logics operation**" & **SI Appendix, Figure S10**). The Boolean logics functions in our device have two key merits. First, all four logics are fully and repeatably programmable in a single 4-node device thanks to the intrinsic self-learning process. Second, the reprogrammable logics are non-volatile since the spatial distribution of magnetic textures is stable after training. Similarly, such reconfigurable Boolean logic can be realized in complex nanostructured networks with nonlinear conductive properties (63). These features allow efficient designs and extensions towards multiple parallel and serial Boolean logics for more complicated and advanced computations.

**Discussion**

We have provided a proof-of-principle study to show that magnetic textures play as a promising platform for self-learning in physical system. In terms of adapting training algorithms, such system offers several potential advantages. First, because the speed of physical evolution is governed by the spin dynamics which is on the order of nanoseconds in ferromagnetic permalloy as illustrated in Figure 5, the training can be accomplished extremely fast without the weight updates from external computers or circuits that usually take much longer time. We note that the relatively long training voltage pulses used (50 μs) in our experiment is due to instrumental limitation of impedance mismatch, and much shorter pulses should be sufficient to train the magnetic texture as shown in our simulation results. Such issues can be potentially solved by optimizing the external circuit connections and employing advanced power source. These efforts will further improve the training efficiency of the network. Second, since the weight updating process is mainly driven by local conductance response to local external stimulus, the self-learning approach is parallel in nature where increased number of nodes would not significantly increase the amount of training time as illustrated in Figure 5 (14, 19, 20). Last but not the least, the stability and high speed of spintronics device mean that they can be rewritten or reconfigured frequently over the lifetime of a circuit, a feature that is essential in many emerging computing concepts (4).

Several concepts related to physical self-learning have been proposed in various systems, offering promising ways to leverage natural phenomena for optimization. In memristor networks, an emergent Lyapunov force allows the system to efficiently explore complex energy landscapes and find optimal solutions (21, 22). A similar approach called "learning by mistakes" has also been proposed in memristor networks, although it still requires peripheral computation for trials and corrections (64, 65). Another interesting concept is equilibrium propagation (24, 66), a learning framework for energy-based models that utilizes two phases (nudged and free phase) to perform learning. This concept has given rise to contrastive local learning, which has been realized in nonlinear resistive networks (67). Additionally, recent work has demonstrated supervised and reinforcement learning in nanowire networks by leveraging inherent nonlinear dynamics and heterogeneous connectivity (68). These approaches, along with physical self-learning realized by magnetic textures, offer promising avenues for allowing nature to perform optimization tasks efficiently.

Although promising as a prototype demonstration, several questions need to be addressed in further studies. One question is how to amplify the output signal since the AMR in ferromagnetic materials are relatively small. A possible way is to use the current In plane (CIP) structure of giant magnetoresistance (GMR) to enhance the conductance difference (46) and the output signals strength can be further boosted by operational amplifiers which is widely used in Complementary Metal Oxide Semiconductor(CMOS) technologies. Scaling such system to maintain good performance requires further validation in future studies. Also, further extension on how to achieve more functionalities, for example, implementations in deep neural network with backpropagation and Boltzmann machine, will be interesting topics to explore in the future.

**Materials and Methods**

**Sample fabrication**

We fabricate three layers: the magnetic layer (Py), the insulating layer ($Al_2O_3$) and the gold layer (Au) as physical self-learning system. The Py layer is deposited at room temperature on high-resistivity $Si/SiO_2$ wafer by DC magnetron sputtering deposition with base pressure of $\leq 2 \times 10^{-7}$ Torr and Ar gas of 2 mTorr. After growth, the sample is annealed at 350° for an hour in ultra-high-vacuum chamber. The sample is etched by ion beam etching (IBE) and then deposited $Al_2O_3$ layer by RF sputtering. Finally, the gold layer is deposited with patterned geometry.

**Electrical setup**

The voltage input in Au layer is supplied by Keysight B2912 source-meter. For electrical measurements in the Py layer, constant source current of 500 uA is applied by Keithley 6221 and voltages are measured by SR830 lock-in amplifier (17 Hz). We use 4-probe technique to remove the contact and wire resistance. The conductance matrix can be obtained by Kirchhoff's law from the measured voltages (see below). The pulse used for training network could be completed by constantly applying multiple 50 μs, 3 V pulses as in Figure 3, or equivalently by applying one single 200 μs, 5 V pulse by Keysight B2902.

**Determining the conductance matrix**

To obtain the conductance matrix of magnetic texture in Py, we use seven wiring geometries(**SI Appendix, Figure S12**). Take Fig. S12a for instance, the $I_1$ and $I_2$ are wired to the positive end of current source (Keithley 6221), while $I_3$ and $I_4$ are wired to the negative end. Respectively, $V_1^{Py}$ and $V_2^{Py}$ are wired to the positive end of SR830, while $V_3^{Py}$ and $V_4^{Py}$ are wired to the negative end and grounded. According to the Kirchhoff's law, the relationship between *I* and *V* can be expressed as:

$I_{\text{input}} = I_1 + I_2 = G_{11} * V_1^{Py} + G_{12} * V_2^{Py} + G_{13} * V_3^{Py} + G_{14} * V_4^{Py} + G_{21} * V_1^{Py} + G_{22} * V_2^{Py} + G_{23} * V_3^{Py} + G_{24} * V_4^{Py}$,

$$V_1^{Py} = V_2^{Py},$$
$$V_3^{Py} = V_4^{Py} = 0,$$
$$V_1^{Py} - V_3^{Py} = V_{measure}$$
$$G_{12} = G_{21}, G_{13} = G_{31}, G_{14} = G_{41},$$
$$G_{11} = -G_{12} - G_{13} - G_{14},$$

In the same way, with seven wiring geometries, we obtained seven equations that contain six independent elements $G_{ij}$ to be solved. Here we use six equations to solve the values of $G_{ij}$. The obtained $G_{ij}$ are inserted in the remaining equation to calculate the voltage, which is within 2% deviation compared to measured voltage.

For the evolution of conductance matrix $G_{ij}$ shown in Figure 3a, the corresponding voltage evolution for seven wiring geometries are measured (**SI Appendix, Figure S13**).

**Number of independent configurations**

For a 4-node binary network, there are 16 ($2^4$) states. In our system, "+ + + +" and "− − − −" are trivial. Due to the mirror symmetry (flipping the voltage sign), there remain seven independent configurations.

**Inference process for associative memory**

The asynchronous inference process is illustrated in the flow chart (**SI Appendix, Figure S14**). For a given voltage pattern (image) and magnetic texture configuration, the following process is employed in the Py layer:

1) Choose a distorted image.
2) Encode voltages $V_i^{Py}$ in the Py layer, wire the electrodes with same polarity and connect to the Keithley 6221.
3) Randomly choose a node (electrode) i.
4) Measure the current $I_{i1}$ flowing through it.
5) Inverse the polarity of electrode i and measure the current $I_{i2}$ flowing through it.
6) Calculate current $I_i = I_{i1} + I_{i2}$.
7) Obtain $V_i^{Py}$ = sgn $(I_i - \xi I_i^{1T})$ [V]. Here, $I_i^{1T}$ is pre-measured by performing steps 3) to 5) when the 1T out-of-plane magnetic field is applied. $\xi$ is set as 0.98.
8) Update $V_i^{Py}$.

Go back to step 2) according to the judgment condition. The inference process stops when $V_i^{\text{Py}}$ remains unchanged for more than 10 iterations.

Compared to asynchronous inference process, the synchronous inference process does not choose a node but renew all nodes simultaneously according to $V_i^{\text{Py}}$. Such process is used in image recognition task shown in Figure 5.

For large 3×5 images, every image is scanned and split into eight 2×2 subimages. Each subimage is encoded by applying voltage pulse in the Au layer and inferred by following the above process. In principle, the eight 2×2 subimages can be encoded by using eight individual magnetic devices and inferred simultaneously. Here as a proof-of-principle, we use one device to perform the above operations independently.

**Micromagnetic simulation**

The intrinsic gradient descent learning is further validated by considering the intrinsic spin dynamics. To illustrate this point, we demonstrate the evolution of the magnetic textures by micromagnetic simulations, which can be performed by the Micromagnetics Module designed for COMSOL Multiphysics. The dynamics of the magnetization $\mathbf{M}(\dot{\mathbf{r}}, t)$ of a ferromagnetic film with saturation magnetization $M_s$ is governed by the Landau-Lifshitz-Gilbert (LLG) equation

$$\frac{\partial \mathbf{m}}{\partial t} = -\gamma \mathbf{m} \times \mathbf{H}_{\text{eff}} + \alpha \mathbf{m} \times \frac{\partial \mathbf{m}}{\partial t}, \tag{S1}$$

where $\mathbf{m} = \mathbf{M}/M_s$, $\gamma$ is the gyromagnetic ratio, and $\alpha$ is the Gilbert damping constant. The effective magnetic field

$$\mathbf{H}_{\text{eff}} = A\nabla^2 \mathbf{m} - K\mathbf{m} \cdot \hat{\mathbf{z}} + H \tag{S2}$$

consists of the exchange interaction, the perpendicular hard-axis anisotropy along $\hat{\mathbf{z}}$, and external magnetic field, parametrized by $A$ and $K$, respectively (see Table 1).

The electric current density $\mathbf{j}$ is proportional to the local electric field $\mathbf{E}$,

$$\mathbf{j}(\mathbf{r}) = \hat{\Sigma}[\mathbf{m}(\mathbf{r})] \cdot \mathbf{E}(\mathbf{r}), \tag{S3}$$

where $\Sigma[\mathbf{m}]$ is the 2 × 2 conductivity matrix of a magnetic thin film with the anisotropic magnetoresistance (AMR)(52), i.e., a local resistivity depending on the angle $\theta$ between the current flow and the local magnetization $\mathbf{m}(\mathbf{r})$ as $\rho = \rho_\parallel \cos^2\theta + \rho_\perp \sin^2\theta$. Inverting this relation, the Cartesian elements $\Sigma_{ij}[\mathbf{m}] = \sigma_\perp + \sigma_\delta m_i m_j$ with $\sigma_\perp = 1/\rho_\perp$, $\sigma_\delta = 1/\rho_\parallel - 1/\rho_\perp$ and the AMR ratio $a = 2(\rho_\parallel - \rho_\perp)/(\rho_\parallel + \rho_\perp)$. Oersted field generating by current distribution in the Au layer can be solved by AC/DC module of COMSOL Multiphysics. Then using Oersted field as input, we solve the spatiotemporal equation S1 and equation S3 self-consistently under the constraint $\nabla \cdot \mathbf{j} = 0$ by the COMSOL Multiphysics finite element code.

**Self-learning magnetic network versus adaptive flow network**

The cost function in our magnetic system (Eq.(3)) is in high resemblance to the cost function theoretically proposed in adaptive flow network(58). Flow network is a physical model where links are strengthened and weakened over time adaptively, aiming to maintain optimal performance under changing loads. Defining a network as a graph of N nodes connected by links $ij$, where $i,j \in \{1,2,\cdots,N\}$ the length of links $l_{ij}$ and corresponding conductance $C_{ij}(t)$ vary over time. Flow rates $Q_{ij}$ between two nodes are linearly dependent on their conductance $C_{ij}$ for fixed potential differences. As is proposed in ref(58), memory can be formed in an adaptive flow network described by the cost function

$$E(t) = \sum_{\langle ij \rangle} \frac{\langle Q_{ij}(t)^2 \rangle_T}{C_{ij}} - \lambda \sum_{\langle ij \rangle} (C_{ij}(t) l_{ij})^\beta l_{ij}. \quad (S4)$$

The first term of Eq. (S4) is exactly the power loss and the second term applies a global constraint in the presence of building cost, with $\lambda$ the Lagrange multiplier and the parameter $\beta$ determines how link conductance contribute to the building cost. The flow network adapts itself by minimizing the cost function Eq. (S4) and is able to encode memory under certain conditions. It should be noted that both cost functions, i.e., Eq. (3) in the main text for magnetic network and Eq. (S4) above for flow network, are equivalent for describing flow network (either electric current of fluid current) governed by Kirchhoff's law and certain physical constraints, while the effective "building cost" in our magnetic device is provided by AMR effect. It is worth noting that in our magnetic network system, the input is encoded as voltages rather than currents. This encoding choice maximizes the Ohmic power loss while simultaneously minimizing the mapped cost function. In contrast, conventional flow network typically employs current as input, leading to minimized power loss when the cost function is reduced.

**Relation between cost function of magnetic texture and energy function of HNN**

In the main text, we experimentally and numerically prove that the evolution of magnetic textures under stimulation follows the minimizing process of the cost function Eq. (3), which can be rearranged as

$$C = \sum_{ij, i \neq j} \frac{\left(V_i^{\text{Au}} - V_j^{\text{Au}}\right)^2 G_{ij}}{2} + \sum_{ij, i \neq j} \alpha_{ij} \left(G_{ij} - G_{ij}^{\max}\right)^2$$

$$= -\sum_{ij} G_{ij} V_i^{\text{Au}} V_j^{\text{Au}} + \sum_{ij, i \neq j} \alpha_{ij} \left(G_{ij} - G_{ij}^{\max}\right)^2$$

$$= -\sum_{ij} w_{ij} V_i^{\text{Au}} V_j^{\text{Au}} - \sum_{ij} G_{ij}^{\text{avg}} V_i^{\text{Au}} V_j^{\text{Au}} + \sum_{ij, i \neq j} \alpha_{ij} \left(w_{ij} + G_{ij}^{\text{avg}} - G_{ij}^{\max}\right)^2, \quad (S5)$$

where we apply the definition $w_{ij} = G_{ij} - G_{ij}^{\text{avg}}$ and the property that $\sum_{j, j \neq i} G_{ij} = -G_{ii}$, arising from current conservation guaranteed by Kichhoff's law, thus

$$\sum_{ij, i \neq j} \frac{\left(V_i^{\text{Au}} - V_j^{\text{Au}}\right)^2 G_{ij}}{2} = \sum_{ij, i \neq j} \frac{G_{ij}(V_i^{\text{Au}} V_i^{\text{Au}} + V_j^{\text{Au}} V_j^{\text{Au}} - 2 V_i^{\text{Au}} V_j^{\text{Au}})}{2} = -\sum_{ij, i \neq j} G_{ij} \left(V_i^{\text{Au}} V_j^{\text{Au}} - V_i^{\text{Au}} V_i^{\text{Au}}\right) = -\sum_{ij} G_{ij} V_i^{\text{Au}} V_j^{\text{Au}}. \quad (S6)$$

During the self-learning process for a given set of $V_i^{\text{Au}}$, the cost function Eq. (S4) is actually being minimized by self-adjusting $w_{ij}$ (note that $G_{ij}^{\text{avg}}$ is constant), leading to modification on its energy landscape. On the other hand, during the inference process of HNN, the network is trying to find a set of $V_i^{\text{Au}}$ for given weights $w_{ij}$ to minimize the energy function Eq. (4). Therefore, the only mutual term that is both function of $w_{ij}$ and $V_i^{\text{Au}}$, i.e., $-\sum_{i,j} w_{ij} V_i^{\text{Au}} V_j^{\text{Au}}$, indicates that the stimulus applied on the magnetic thin film (voltage on each Au electrode) has been encoded (or memorized) as the ground state of the HNN according to the physical self-learning process.

**Boolean logics operation**

To operate as Boolean logics, two nodes are treated as inputs (node 2 and 3), one as output (node 4) and the remaining one as control (node 1) (**SI Appendix, Figure S10**). Hence, the magnetic device plays as a 2-input control gate. We use learned pattern "$+ - - -$" to achieve AND (OR) gates when the control input is 1 (0). Experimentally, in the Py layer, we apply voltages on control and input nodes and the output signal can be obtained according to $\text{sgn}(V_2^{\text{Py}} * w_{24} + V_3^{\text{Py}} * w_{34} + V_1^{\text{Py}} * w_{14})$ from node 4 (**SI Appendix, Figure S11**). Function can be reprogrammed, e.g., from AND to OR by flipping the voltage applied on the control node 1, whose contribution to the output $V_1^{\text{Py}} * w_{14}$ is equivalent to a bias which determines the threshold of final output. Remarkably, the magnetic texture can be reconfigured with input pattern "$+ - - +$" for logic NOR (NAND) gates, which are operated in the same way as AND (OR).

**Acknowledgments**

This work was supported by the National Key Research Program of China (2022YFA1403300, 2020YFA0309100), the Innovation Program for Quantum Science and Technology (Grant No.2024ZD0300103), the National Natural Science Foundation of China (11991060, 11974079, 12074075, 12074073, 12074071, 12204107, 12274083), Shanghai Municipal Science and Technology Major Project (2019SHZDZX01), Shanghai Municipal Natural Science Foundation (22ZR1407400, 22ZR1408100, 23ZR1407200), Shanghai Science and Technology Committee (21JC1406200), and Pujiang Program (21PJ1401500). Part of the experimental work was carried out in the Fudan Nanofabrication Laboratory. All authors are grateful for the fruitful discussions with Prof. Zhe Yuan.

**Figures**

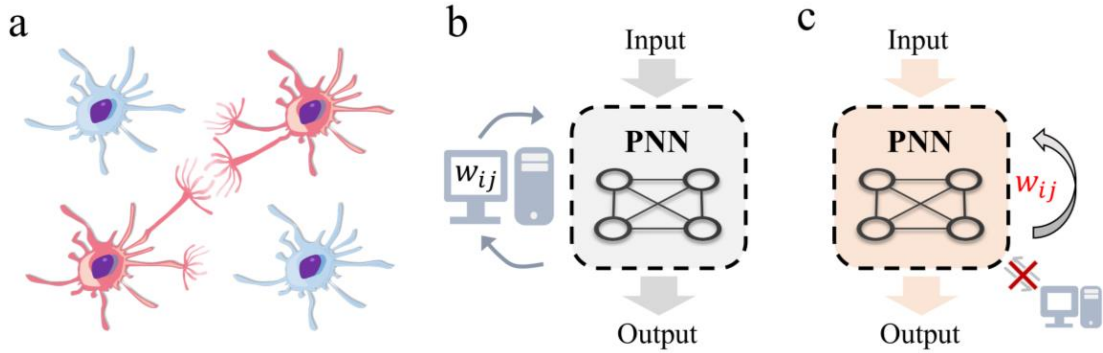

**Figure 1. Concept of physical self-learning.** a) Biological neural network learns by itself where activated neurons intend to strengthen their connection. b) Physical Neural Networks (PNN) whose internal learning parameters $w_{ij}$ are determined via external computation and updated physically. c) PNNs with self-learning capability, whose learning parameters are determined and updated in an autonomous way according to inherent physical dynamics, without interference of external computation.

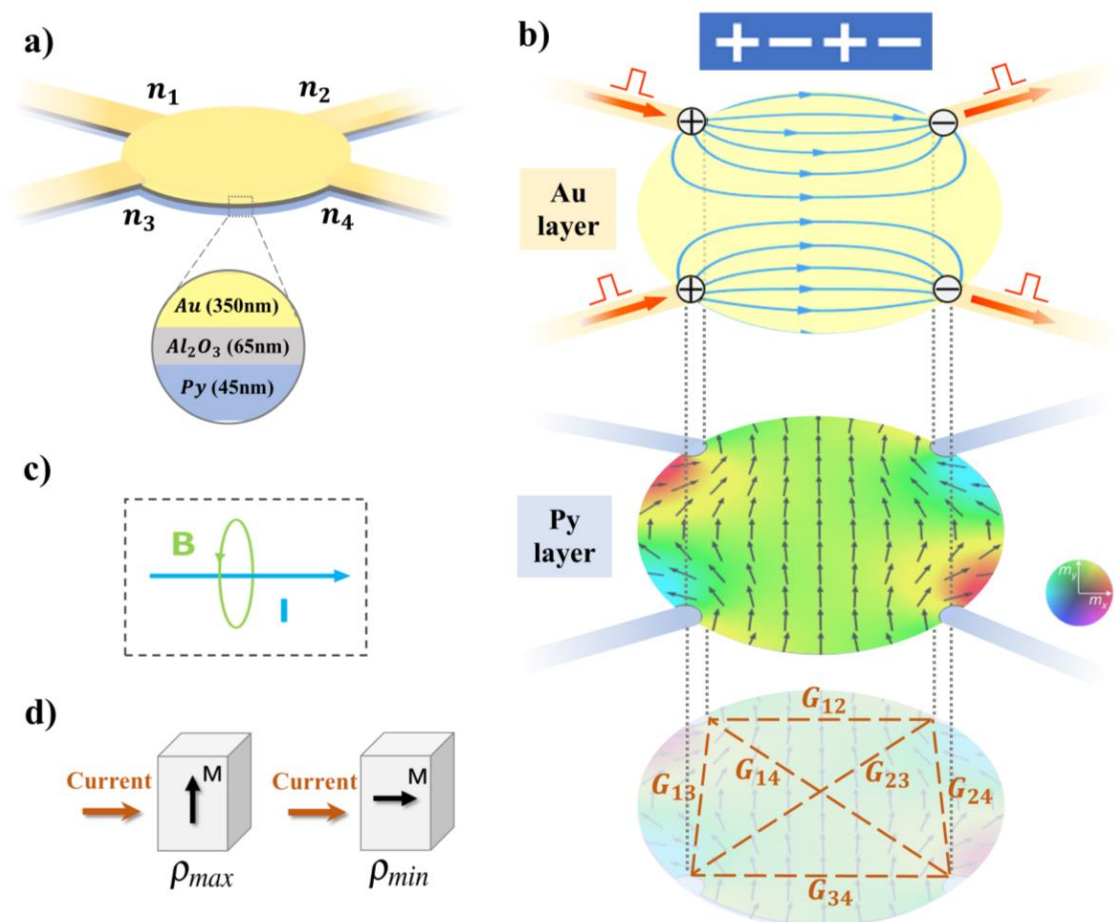

**Figure 2. Working principle of a 4-node self-learning magnetic HNN.** a) Schematic of experimental device structure. b) demonstration on the working principle of magnetic HNN. The top Au layer serves as neurons to input data. For pattern "$+-+-$", the input voltage pulses generate a current distribution in the Au layer (top) and configure the spin orientation in the Py layer (middle) via Oersted field effect in (c). The evolution of spin orientation (i.e. magnetic texture) can be described by the conductance matrix $G_{ij}$ (bottom) which mimics the synaptic weights in HNN. c) The Oersted field effect. d) The AMR effect in Py layer is employed to detect $G_{ij}$ evolutions where the resistance depends on the angle between the detection current and the magnetization direction.

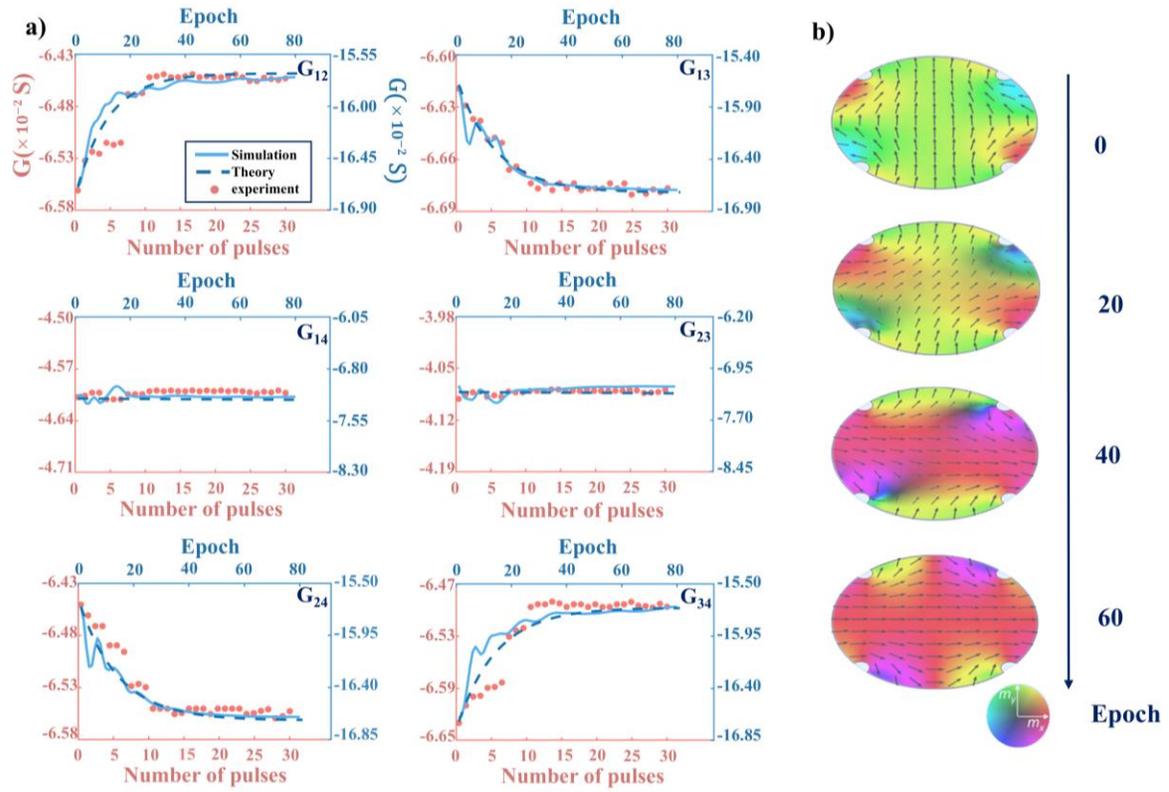

**Figure 3. Intrinsic gradient descent learning.** a) Evolution of conductance matrix $G_{ij}$ when the voltage pattern is switched from initial "+ − + −" to "+ + − −" under "+ + − −" voltage pulses. The experiment, simulation and theory curves are plotted respectively where the matrix evolution is equivalent to the Oja's learning rule for unsupervised learning. b) Snapshots of corresponding evolution of spin textures from simulation.

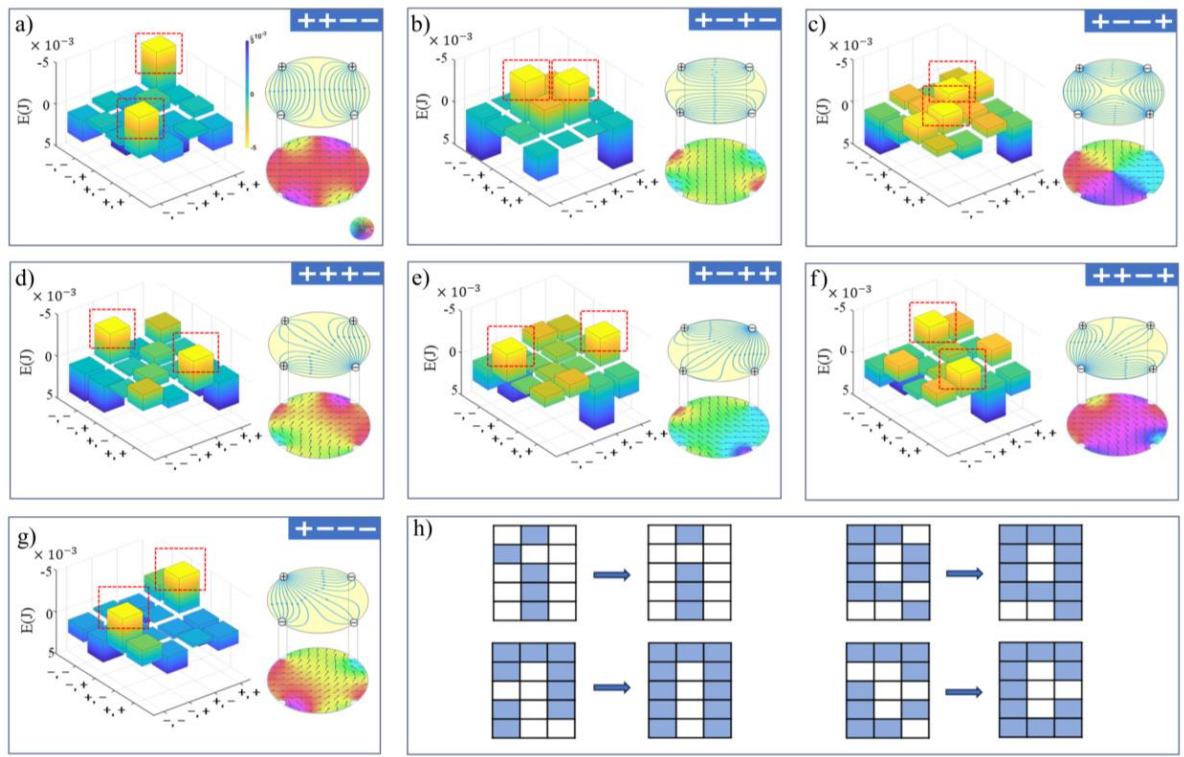

**Figure 4. Examination of the magnetic HNN.** a) The effective energy diagram (left) and corresponding current and spin texture distribution (right) for the trained pattern "+ + − −". The effective energy $E$ is minimal for the trail state "+ + − −" or equivalent "− − + +" as marked by red dashed box. b)-g) Results for other six trained patterns. h) Associated memory. When a distorted letter is fed to the network, the correct letter can be recalled during inference. Four letters (i, n, q, c) are demonstrated respectively.

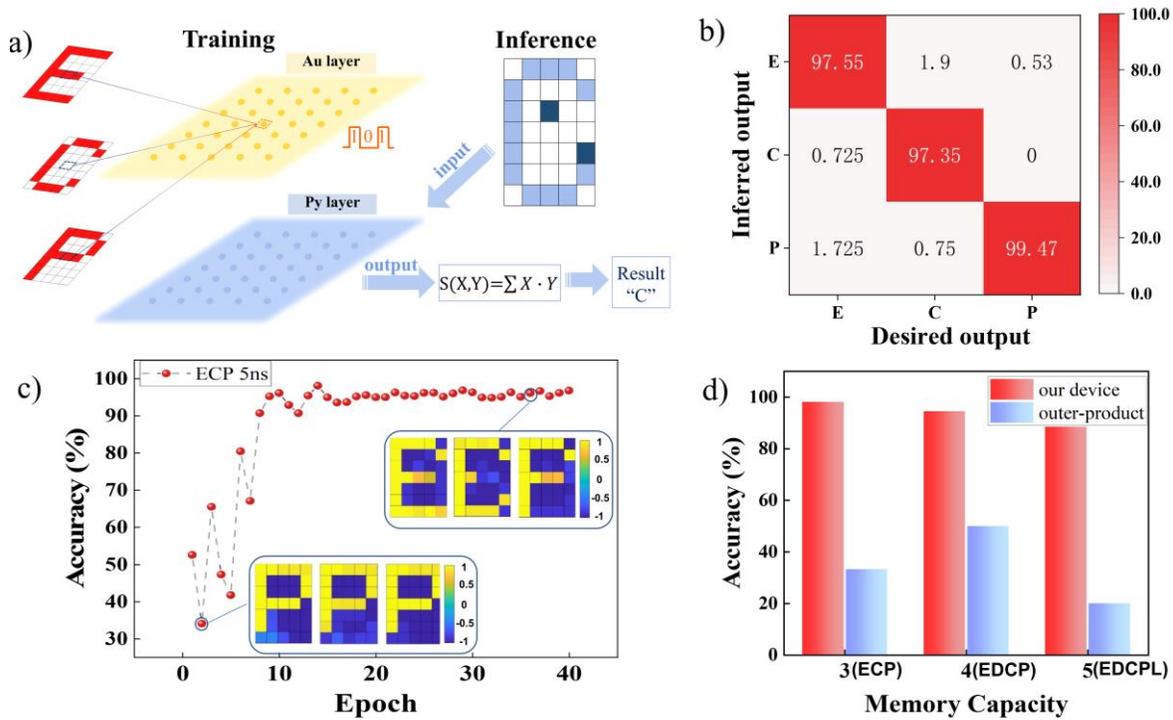

**Figure 5. Associative memory of similar patterns with a 35-nodes magnetic HNN.** a) The process flow. The three letters (E,C,P) is first converted into pulse of different nodes and fed to Au layer orderly. Each letter is trained for 5 ns. After training, a distorted pattern is fed to the Py layer (with low voltage). After synchronous inferring, we calculate *S(X,Y)* and the corresponding output. b) Confusion matrix shows the obtained accuracy from the system vs. the desired outputs. The occurrence of the predicted output for each testing set is represented by colors shown in the color scale. c) Evolution of the accuracy with iterations. After about 9 epochs, our device can get about 95% accuracy. d) The recalling accuracy of our device and outer-product training rule with different memory capacity. The results are produced by micromagnetic simulations.

**Table 1. Parameters for micromagnetic simulation** (69)

| Name | Parameters | Value | Dimension |
|---|---|---|---|
| **Exchange constant** | $A$ | 2.5863e-11 | A·m |
| **Hard-axis anisotropy** | $K$ | -43.062 | A/m |
| **Saturation magnetization** | $M_s$ | 8e5 | A/m |
| **Gilbert damping** | $\alpha$ | 0.3 | 1 |
| **Gyromagnetic ratio** | $\gamma$ | 2.21e5 | Hz/(A/m) |
| **Isotropic conductivity** | $\sigma_0 = \sigma_\perp + \sigma_\delta/3$ | 5e6 | S/m |